\documentclass[12pt,a4paper,aps]{article}
\usepackage{a4}
\usepackage{epsfig}
\usepackage[hmargin=2cm,vmargin=2cm,nohead]{geometry}
\begin{document}

%\begin{center}
%{\huge \bf 
\title{The Future of Nuclear Energy: Facts and Fiction \\ 
An update using 2009/2010 Data} 
%\end{center}

\author{
Michael Dittmar\thanks{e-mail:Michael.Dittmar@cern.ch},\\
Institute of Particle Physics,\\ 
ETH, 8093 Zurich, Switzerland\\
\date{January 21, 2011}
}
\maketitle
%\par
%\begin{center}
%{\large \bf 
%\author{Michael Dittmar \\ July,  2009}
%\end{center}

\begin{abstract}
%The fog, which hides a clear view about the future use of energy, has become more transparent 
%during the first decade of the 21 century. However, more and more people observe that instead of a blue and sunny sky 
%thunderstorm clouds are approaching the rich energy intense 
%industrial ``way of living". Some observers start praying for techno-fix alternative energies, and especially for the nuclear energy option.  
%Others conclude that this energy intense industrialized culture will not survive the approaching storm.

An update of our 2009 study, ``The Future of Nuclear Energy, Facts and Fiction" using the 2009 and the available 2010 data, including a critical look 
at the just published 2009 edition of the Red Book,  is presented.  

Since January 2009, eight reactors with a capacity of 4.9 GWe have been connected to the electric grid and 
four older reactors, with a combined capacity of 2.64 GWe have been terminated.
Furthermore, 27 reactor constructions, dominated by China (18) and Russia (4), have been initiated. 

The nuclear fission produced electric energy in 2009 followed the slow decline, observed since 2007, with a total production of 2560 TWhe,
41 TWhe (1.6\%) less than in 2008 and roughly 100 TWhe less than in the record year 2006. 
The preliminary data from the first 10 months of 2010 in the OECD countries indicate that 
nuclear power production in North-America remained at the 2009 levels, while one observes a recovery in Europe with an increase of 2.5\% and 
a strong rise of 5\% in the OECD Asia-Pacific area compared to the same period in 2009. 

Worldwide uranium mining has increased during 2009 by about 7000 tons to almost 51000 tons. 
This increase is dominated by Kazakhstan, the new uranium ``Saudi-Arabia",   
which reported an increase of about 6000 tons (63\%) to 14000 tons.  
Still roughly 18000 tons of the 2010 world uranium requirements
need to be provided from the civilian and military reserves. 

Perhaps the most remarkable new data from the just published 2009 edition of the Red Book, 
are that (1) the best understood RAR (reasonable assured) and IR (inferred) resources, with a price tag of less than 40 US dollars/Kg, have been 
inconsistently absorbed in the two to three times higher price categories and 
(2) uranium mining in Kazakhstan is presented with a short lifetime. The presented mining capacity numbers indicate an uranium extraction peak of 
28000 tons during the years 2015-2020, from which it will decline quickly to 14000 tons by 2025 and to only 5000-6000 tons by 2035. 
\end{abstract}

%\maketitle

\newpage
\section{Introduction}

%Discussions about alternatives to todays oil, gas and coal addictions  
%become quickly highly emotional, especially when the situation with the future of   
%nuclear fission and fusion energy is being discussed. 

The worries about existing and potential problems with our oil, gas and coal supplies for our 
industrial based way of living have certainly increased during 2009 and 2010 \cite{worries}.
However, people enthusiastic about large scale technology, especially with some background in physics, 
are pointing quickly either to nuclear energy or to large scale solar power projects as possible solutions to such worries\cite{scienceart}. Such views are supported by 
most economists and politicians who propose that one only needs to invest trillions of dollars to manage potentially existing problems with our fossil 
fuel based energy civilization \cite{economicview}. It is assumed that the intelligent investment of money will be sufficient to solve the problems 
with nuclear fusion and make it available either directly on our planet or indirectly using solar energy. 
Such views about the future use of energy are often lacking the relevant facts about today's energy use 
and existing technological constraints. Instead, the preferred form of discussions seems to be dominated by  
theoretical and hypothetical ideas about unproven concepts with unknown capital costs. 

The idea of our review ``The Future Of Nuclear Energy, Facts and Fiction", presented in four chapters during the second half of 2009 
to the Oil Drum community, \cite{oildrumart}, was to provide reliable background material for such discussions. 
It might be argued that the information presented in these ``Oil Drum" articles reached, somewhat surprisingly,
attention far beyond the Oil Drum readers.  Some arguments from these papers where even picked up by major newspapers
and more popular scientific journals like ``New Scientist" and ``Scientific American" \cite{mediareaction}.   

To be sure, and perhaps as should have been expected, the hard numbers regarding the underlying and often hidden nuclear reality that were  presented 
in that review were like a red flag for believers in a bright future of nuclear energy.  Two bets about the yearly evolution of the number of produced TWhe of nuclear energy
and about the yearly uranium extraction resulted from these discussions \cite{nuclearbets}. 
 
Concerning the first bet, the number of produced electric energy in 2009 was 41 TWhe (1.6\%) smaller than in 2008,
corresponding roughly to the ``shutdown" of nuclear power plants with an equivalent of  6 GWe. 
The preliminary data from the first 10 months of 2010 in the OECD countries indicate that the decline in the OECD has been stopped and that one can expect now that the 2010 results 
will be roughly equal to the 2008 results \cite{IEAdata}. 
%about the and that North-America of 2\% compensated by an increase in Europe of 2.1\% 
%and a rise of 4.7\% in the OECD Asia-Pacific area compared to the same period during 2009 

While the observed decline of 1.6\% for 2009 was stronger than my estimated average yearly 
decline of 1\% during the period 2009/2010 the recovery observed so far during 2010 are 
thus indicating that the yearly fluctuations, at least for now, are stronger than the effects 
expected from the new power plants and from the overall aging of nuclear power plants.     
More details about the actual situation will be given in section 2.1.   

Concerning the second bet, the 2009 uranium mining results of 51000 tons exceeded my 
guesstimate of 45000 - 47000 tons by about 6000 tons\footnote{This 2009 result matches well with the plans presented during the last years from Kazakhstan, 
which I had assumed to be unrealistic.}\cite{mining2009}. However, this increase comes essentially only from the reported Kazakhstan results and the extraction from 
all other countries, despite a relatively high price for uranium,  remains in agreement with my expectations.

First estimates and results from 2010 indicate a similar trend for the year 2010 with an expected result of 55000 tons. Again, this will come only 
from Kazakhstan with an estimated increase from 14000 tons in 2009 to 18000 tons for this year\cite{mining2010}.
Further large increases from this country alone are already announced and 
uranium mining is claimed to grow further to even 30000 tons by 2018. 
More details about the latest developments of uranium mining will be presented in section 2.2 and 3.1. 
The next few years will show if uranium mining in Kazakhstan can continue to grow and if 
my view about significant natural limitations for future production of fission energy originating from uranium mining requires some revisions.   

The most remarkable evolution concerning future nuclear technologies is the apparent international loss of support for ITER, the 
world thermonuclear plasma fusion project. Critical arguments where presented in a Scientific American article, 
``Fusion the false dawn" in March 2010 translated at least to its German and French editions\cite{scientificamerican}.
This loss of support reached almost dramatic levels during the past 6 months as original estimates 
of construction costs were revised upwards by almost a factor of three\cite{ITERcost}.
More ``news" about the fission and fusion research will be presented in section 2.3 and 2.4 

In section 3 we describe the situation and analyze prospects with nuclear energy 
in the different OECD blocks, Europe, North-America and Asia and for Russia, China and India 
the larger nuclear energy users outside of the OECD block.  
 
Finally, in section 4, we present some of the most  
amazing uranium resource changes reported in the just published 2009 edition of the Red Book \cite{RB09}. The bi-annual Red Book,
is a joint document from the IAEA (International Atomic Energy Agency from the UN) and the NEA (Nuclear Energy Agency 
from the the OECD countries) which has provided for more than 40 years an official data base for 
the estimated exploitable uranium resources. 

In order to make this update readable without reference to the original article~\cite{oildrumart}, a short summary of the basic facts about 
nuclear energy is given in section 1.1.

\subsection{Using nuclear fission energy, some basic facts}

Nuclear energy is released by the neutron induced fission of the uranium isotope U235, 0.71\% of natural uranium,    
and the plutonium isotope Pu239.
The released energy per nuclear fission reaction is about a hundred million times larger than in any chemical molecular reaction. 
The released fission energy is carried by the fission products and is then transferred to water molecules by elastic collisions within the reactor. 
The resulting heated water is used, similar to fossil fuel power stations, to produce electric energy via the Carnot cycle.

Existing nuclear power plants transform only about 33\% of the total produced fission energy into electric energy and the rest is almost entirely 
lost as waste heat. The reason for this relatively low efficiency, compared to modern coal and gas power plants, comes from the lower 
operational temperature, chosen among other reasons to limit corrosion effects in the difficult to replace pressure tank. 
Because of (a) the remote location of nuclear power plants and (b) the relatively low steam temperature,  
the waste heat is rather unattractive for secondary uses. 

Commercially used nuclear power plants are currently operated in 30 countries. 
Now there are 442 reactors with a total electric power capacity of 374.9 GWe (January 2011) connected 
to the electric grid. In 2009 the reactors produced a total of 
2560 TWhe, about 41 TWhe smaller than in 2008, electric energy. 
The preliminary data for 2010 indicate that a total production similar 
to the one in 2008 can be achieved. Taking roughly the average between 2009 and 2010 one finds that this number is roughly 70 TWhe lower than during the nuclear fission record year of 2006.  
Perhaps one could use the ups and downs in the years 2007 to 2010 of the nuclear produced electric energy to constrain the yearly random fluctuations due to unpredictable outages 
in the nuclear power operation to $\pm 1\%$ or $\pm 25$ TWhe/year.  
In comparison, hydropower produced electric energy has increased continuously during the past years 
to more than 3000 TWhe in 2009 \cite{IEA,PRIS}.  

In terms of its total contribution to energy consumption, only about 14\% of the world's electric energy comes from nuclear fission (down from 18\% some 15 years ago) and electric energy itself is only 
about 16\% of all energy consumed. 

When thinking about a large increase in nuclear energy, it is important to consider that  
nuclear reactors work most efficiently when operated in a stable mode at 100\% power capacity.  
Nuclear power plants thus provide an option for the base load of regional electric grid systems. 
One consequence of the daily and yearly demand variations for electric energy is that 
countries which produce a large fraction of their electric energy from nuclear power have a nightly overproduction.
This overproduction is often used for simple electric heating systems and ``luxury" applications. 

Another consequence of the nightly overproduction from nuclear power plants are 
large geographic, often ``multi-country", electric grid systems. Such systems are usually backed up with hydropower pump storage systems, 
which can be ``recharged" using the nightly overproduction. 
Such existing grids manage to integrate nuclear produced electric energy fractions between roughly 20 to 40\%. 
Perhaps the best example of a large multi country electric grid system is found in Western Europe. 
This european grid functions efficiently with a total of 26\% of its electric energy from nuclear power plants, integrated 
into a sophisticated large-scale system of hydropower pump stations in its mountainous regions. It is remarkable that France, with the highest fraction of nuclear power of any country in the world, has became a net importer of electric energy during peak load winter times\cite{peakloadstress}.  

Generally accepted solutions for a safe storage of radioactive waste, despite expensive attempts in many countries,  have so far not been found.
And claims that the retail price of nuclear produced electric energy around the world is high enough to pay for appropriate 
decommissioning of the nuclear power plants and safe storage of nuclear waste are at best a 
half truth\cite{decommissionfund}.
While it is correct that many countries require that a fraction of the electric energy bill is put aside for the later decommissioning and 
the final waste storage, it is impossible to estimate accurately the real final costs and it is unclear how  
inflation is taken into account. In addition, the estimated amount of money will be accumulated only over the estimated entire future lifetime of the nuclear power plants.
Thus, even in the unlikely case that waste solutions can be found that cost no more than has been estimated and allowed for, future generations will always be confronted 
with unplanned early terminations of nuclear power plants and the resulting huge financial ``black holes". 

In addition, even though multi billion dollars have already been put aside in many countries, little transparency about these funds exists.
One can guess however that this money is currently used for (a)``research and technological development" of 
a final nuclear waste storage system with some comfortable well payed jobs and sometimes even for (b) ``safe" financial speculations \cite{speculations}
to ``help keep up with inflation". 

Pressurized water reactors, the dominant form of existing and planned nuclear power plants 
require U235 (or Pu239) enriched to 3-4\%, corresponding to  
roughly 170 tons/GWe/year of natural uranium equivalent and about 500 tons for their first uranium fuel load. It follows that 
the startup for a large number of new reactors requires a substantial extra amount of uranium and that some 1-2 year fuel reserves are 
stored in running reactors. One can imagine that under ideal planning, 
this extra nuclear fuel can be used during the termination period of old reactors.   

For at least 15 years, only about 2/3 of the uranium fuel is coming from current mining, the other 1/3 coming from secondary sources, 
like civilian and military stocks. It is generally agreed that the civilian secondary stocks will essentially be exhausted 
during the next  5 to 10 years and will have to be replaced by newly uranium mining \cite{uraniumneeds}. 
The possibilities to increase uranium mining from about 50000 tons during 2009 to 70000-100000 tons during the coming decade depends on many assumptions and should be a matter of heated debate. 

Believers in a bright future of nuclear energy usually argue either: 
\begin{itemize}
\item[1] that an almost infinite amount 
of uranium is in principle available, if one is only willing to pay for higher exploitation 
cost; that the uranium price is currently only a relatively small contribution to the overall cost of nuclear 
produced electric energy; and hence that uranium supply will not be a problem for decades or even centuries.
\item[2] that the amount of exploitable uranium is indeed limited, 
but the almost-mastered nuclear fission breeder technology and/or the use of the Th232 (thorium) to
U233 breeding cycle will allow to 
avoid uranium shortages and will lead to essentially unlimited fissile fuel resources.
\end{itemize}

Critics counter those arguments with the fact that many countries, which rely on nuclear produced electric energy, 
have essentially ceased to extract uranium in sizable quantities from uranium mines located in their own territories.   
The critics argue further that neither the fast breeder technology nor thorium breeder reactors currently exist and that 
sensible political decisions of today should not be based on uncertain and ``yet unknown" technological breakthroughs.

Within this context it is remarkable that the nuclear energy lobby and the corresponding newspaper articles almost never mention how much uranium is required 
to operate the existing 442 nuclear energy reactors and that during past years only about 2/3 of this uranium came from mines. They also fail to mention that 
even the hoped for modest 1\% annual growth of nuclear power capacity requires a huge effort to keep existing mines productive and to open many new uranium mines during the next decades \cite{examplelobby}. 

In summary, the debate about the future role of nuclear energy will continue for many more years until 
the fog from the  remaining ``unknowns" becomes thin enough to distinguish wishful thinking from reality.
Despite its relative small contribution to our overall energy consumption it might thus be interesting to follow the evolution of nuclear energy closely and especially to 
compare past predictions (hypothesis) with the now known subsequent realities (experiment).  
Perhaps this method can lead to improved judgements about nuclear energy and its future limitations.

\section{The latest evolution of nuclear energy: \\
News from the years 2009 and 2010.} 

The evolution of commercial nuclear fission and fusion energy has been plagued from its beginning, more than 50 years 
ago, with unrealistic and false promises. For example: 
\begin{itemize} 
\item nuclear energy growth scenarios from the last decades were almost never realized; 
\item reactors were rarely built within the estimated budget and time scale;
\item the nuclear waste problem is still unsolved and thus left for future generations; 
\item statements that nuclear generated electricity is cheaper than 
other options, producing little or no CO$_{2}$, depend on many hidden assumptions -- for example on 
how the gray energy for construction and for uranium mining is accounted for and how the long term costs for the next couple of decades and even centuries are estimated; 
\item commercial nuclear fusion energy is still "just around the corner" as it has been for decades.
\end{itemize}

Despite this historical legacy of errors, believers in a future  ``nuclear paradise" 
have a long list of explanations of why past promises were not realized. They usually blame 
incapable politicians, ``greenies" and others for past failures. 

It follows that {\it ``from now on"}, everything will be different and that all promises will work according to plans.
Since the meaning of  {\it ``from now on"} will certainly remain rather flexible,  
we propose to continue the comparison of past and present estimates from the NEA/IAEA Red Book and from the World Nuclear Association  
with the real evolution of nuclear energy use during the past few years. 

\subsection{Nuclear Energy: News from 2009 and 2010}

\subsubsection{New reactors}

According to the IAEA database\cite{PRIS}, eight nuclear fission reactors with a capacity of 4.9 GWe 
have been connected to the grid since 2009. Two started 
in 2009, another two in March, one in July and two in August 2010 and one in January 2011. 
During the same period, four older nuclear reactors with a capacity of 2.6 GWe have been terminated --
two in Japan, one in France (the 0.13 GWe Phenix breeder prototype reactor) and one
in Lithuania. The 1.185 GWe Ignalina 2 reactor in Lithuania 
terminated its last operational year with a record production of 10.025 TWhe.

As a result, the net nuclear capacity in Japan has decreased by 0.5 GWe but    
increased by 0.95 GWe in Russia, by 0.6 GWe in India, by 1.6 GWe in China and by 0.96 GWe in South-Korea. 
According to the WNA estimates from September 2010\cite{WNAnewreactors}, using  
official past and present construction schedules, another 4 reactors (3 GWe) should have started 
during the last months of 2010. However, more accurate schedules can be found in the more 
detailed WNA country information papers. These papers indicated already in September 2010 
that the start of these 4 reactors was already moved to 2011. 

According to the PRIS data base a country-by-country analysis during the past decade indicates that on average three 
reactors per year could be connected to the grid and a similar number of older reactors were terminated. The 2008-2010 period 
with 7 new reactor connections (one small Indian reactor was connected to the grid in January 2011) 
looks like any other period during the past 15-20 years.  

However, this situation is supposed to change drastically during the coming years and, as specified in the January 2011 WNA document, 13 reactors should start in 2011~\footnote{According to the more detailed 
country data base of theWNA, only two of these 13 reactors are expected to start early in 2011, while the start of the others is scheduled either for the last months of 2011 or is marked as uncertain.}, 13 reactors in 2012 and similar numbers in the following years. 
It might be interesting that the corresponding WNA document from February 
2007 specified that from the five reactors connected during 2010, only the two in China and the one in Korea were on schedule, 
while the other two were scheduled to start during 2008 (India) and 2009 (Russia). 
Furthermore, according to the 2007 version of the WNA document\cite{WNAnewreactors}, another five 
of the reactors scheduled to start in 2011 should be running already, one in 2007, two in 2008 and two in 2009. 

Since some of the 50-100 old reactors, operated for 30-40 years, will be terminated during these years, it remains to be seen if a net nuclear power increase 
of up to about 7 GWe/year can be realized during the coming decade. 
  
\subsubsection{Electricity from nuclear fission power in 2009}

The number of produced TWhe of electric energy has decreased further by another 41 TWhe 
to 2560 TWhe, about 1.6\% lower than during the year 2008\footnote{Slightly below my last summer guesstimate of 2575 TWhe for 2009, or -1\%/year, 
for the next few years.}. The 2009 result is almost 100 TWhe smaller than the 2658 TWhe achieved during 
the record year 2006.
For 2010, the OECD countries, with about 80\% of the total nuclear capacity, have already reported results for the first 10 months. 
These preliminary 2010 data indicate a stabilization in many OECD countries and 
even an overall increase of 1.9\% (+34 TWhe) for the first ten months compared to the same period of 2009. The reported strongest increases come from Japan (+8.1\%), where a few reactors have restarted after some longer stops, from France (+3.1\%) and from Germany (+4.8\%). 

Another already known ``loss" of 10 TWhe  
with respect to 2009 comes from the terminated Ignalia 2 reactor in Lithuania
which had a record production during its final year. 

As it usually takes at least a couple of months before a newly grid connected reactors reach nominal capacity,
it seems rather impossible that the additional reactors will lead to a significant increase of nuclear produced 
electric energy already during the year 2010.

It is interesting to notice that this yearly decline of about 0.5-1\% per year, observed since the record year of 2006, coincides with the beginning of the so-called ``nuclear renaissance" phase. 
One might thus argue that the nuclear renaissance will ``probably" start only during the next year(s). 

\subsection{Uranium mining 2009 and announcements for 2010.} 

The official worldwide uranium mining results for 2009 are 50772 tons, about 15\% larger 
than in 2008\footnote{Thus about 6000 tons larger than my own guess from August 2009. Assuming that the Kazakhstan numbers 
are correct, I clearly lost the associated bet.}, covering about 76\% of the 2010 uranium requirement\cite{mining2009}. 
This 2009 mining result agrees well with the ones predicted by the corresponding WNA estimates from 2008. 

This increase is based almost entirely on the results reported from Kazakhstan, which increased the annual production, 
in almost perfect agreement with the past official plans, from 8521 tons to 14020 tons. Uranium production in Kazakhstan comes 
from many relatively small mines with a production capacity between 500 tons to 2000 tons.  However, assuming that the 
numbers from the WNA document about the 26 largest uranium mines with an output of more than 500 tons are correct, 
only about 10500 tons of the 14000 tons from Kazakhstan are accounted for in their 10 larger mines\cite{KazakhstanWNA}. 
Some transparency about the origin of the remaining 3500 tons would certainly be welcomed by many observers.  

Uranium mines in Canada have recovered from some past years problems and 
reached again the 2006 levels of about 10 000 tons. 

Despite this impressive increase in Kazakhstan, it is remarkable that the reported uranium mining from all other countries 
has stagnated for several years at around 37 000 tons.  
This is rather surprising if one takes into account that the uranium spot price (for U3O8) has increased from 
10-15 dollars per pound before 2004 to about 68 dollars/pound at the time of this writing\cite{uranspotprice}. 
 
The 2010 expected increase to 55000 tons will again come almost entirely from Kazakhstan, with a planned increase 
from 14000 tons to 18000 tons. According to the latest announcements 
by the Kazatomprom from the 9th of August this year\cite{Kazatomprom2010}, the mining proceeds again perfectly with respect to past years plans: 
 
{\it ``Uranium production volume in the Republic of Kazakhstan for the first six months of 2010 made up 8,452 tU that is 42\% 
more than in the corresponding period of 2009. According to the plans corrected uranium production in the second half of 2010 is expected to be 9,770 tons."}

The prospects for new larger uranium mines in Canada, Australia, Niger and Namibia are less clear.
For the new mines in Canada, the latest announcements say that the start up of the Cigar Lake mine, designed for a yearly production of 7000 tons/year
has been further delayed to sometime in the year 2013 and full production might be reached at earliest during 2016.
The second large project, the Midwest, originally scheduled to start in 2011, has been postponed for several reasons including a large rise in the initial capital cost.
Recent news about large mining projects in Australia, Namibia, Niger and Russia are also rather mixed and some delays 
have already been announced\cite{WNAnewmines}.

Under the assumption that an additional capacity of roughly 50 GWe needs to be fueled with uranium by 2015 and that 
the conversion and delivery of military uranium from Russia will stop at the end of 2013, up to 70 000 tons of uranium, about 20 000 tons 
more than in 2009, need to be mined by 2015. According to the WNA estimates, about half of the required increase
will come from Kazakhstan which plans to reach 24000 tons by 2015.
The other 10 000 tons are predicted to come mainly from new big mines currently under construction in Namibia and Niger.  

In summary it remains to be seen if existing and future uranium mines will be able to increase production from 51000 tons in 2009 to 
about 70000 tons by 2015, required to fuel today's 374.9 GWe plus an expected additional nuclear capacity of about 50 GWe.

\subsection{New types of nuclear reactors}

The french version of the Generation III reactor technology,  
the 1.6 GWe EPR reactor from AREVA, currently under construction in Finland and in France 
has not only seen cost explosions and delays but is now confronted with severe design problems\cite{AREVAproblems}. 
A possible consequence of these difficulties was the announcement that the three planned reactors in the Arab Emirates 
will be constructed by a Korean company and not by AREVA\cite{Arabkorea}.
Similar negotiations with countries like the UK and the USA are currently blocked by considerations about safety and financing\cite{AREVAproblems}. 
Recent news from AREVA seem to indicate a major change, away from their sinking ``1.6 GWe EPR flagship", to a new smaller reactor design\cite{newEPR}.

Fast reactors are operated with the prompt neutrons from the nuclear fission and  
can be considered as prototypes for future Generation IV reactors. 
Among them is the French Phenix reactor, originally a 200 MWe prototype breeder reactor which started operation in 1974. This 
reactor was finally closed in October 2009 and officially terminated in February 2010. One might hope that a detailed review 
of its past operation, allowing verification of the so-far unsubstantiated claims about nuclear fuel breeding, will be published eventually. 
Another fast reactor is the   
Russian BN 600,  a 0.56 GWe reactor, operated since 1981. This reactor has just received the permission to continue its operation for another 10 years.
The construction of its new larger version, the BN 800, was started during 2006 and following some delays 
is currently expected to begin operation in 2014 at earliest.
According to the WNA document,  both Russian reactors are now 
called ``fast reactors" instead of ``fast breeder reactors" \cite{WNArussia}.

The last ``just functioning" fast breeder is now the Japanese Monju reactor with a capacity of 280 MWe. 
It has, after a 15-year-long stop,  officially begun to operate under test conditions this spring and is supposed to 
reach its normal operation only by 2014. 

According to the Japanese nuclear agency the date for the operation of commercial fast breeders 
is now given as 2050\cite{japanbreeder}. This is not very different 
from statements of French government officials claiming that the next commercial 
breeder reactor will not be operational before the year 2040\cite{frenchbreeder}. Thus, the original timeline of 2030 for the Generation IV reactors  
appears to be totally outdated. 

Not much experimentation with respect to thorium reactors has been reported during the past year\cite{thoriumforum}. However, the program to develop a high temperature high efficiency 
pebble bed thorium reactor project in South-Africa has been stopped\cite{pebblebed}. It follows that 
those who hope for future TH232 breeder reactors are left with nothing but modeling to support their hopes. 
Taking the current financial world crisis into account, the construction and realization of 
thorium prototype reactors during the coming years does not look very promising. 

Finally, newspaper reports about future "small" scale wonder nuclear reactors appeared during the past months, claiming that  
such projects are supported by some rich private promoters and also by 
the US and Russian governments\cite{smallreactors}. However, looking at similar claims and plans from past decades one might 
give them not much more credibility than most people give to snake oil medicine. 
%  http://en.wikipedia.org/wiki/Snake_oil

\subsection{Difficulties with the ITER project}

The 2009 and 2010 news about the ITER plasma physics project, known also as the path to commercial nuclear fusion energy, 
a multi billion dollar/euro dream project of all larger countries, demonstrates that it is becoming
nothing short of a financial nightmare for high level powerful bureaucrats and politicians in Brussels and elsewhere\cite{iterfire}. 

Reasons for this nightmare are many:

\begin{itemize}
\item The exploding budgetary costs, now admitted to have increased, even before real construction has started,  by almost a factor of three.
Claims are made, and even transmitted by project officials that ITER became so expensive because it has been set up as an international project.
Certainly not a good start for a project defined to pave the path towards fusion energy.
\item (Doubtful) Statements by scientists and politicians that alternative and much more promising directions need to be taken towards mastering nuclear fusion.  
Examples are the multi billion euro laser fusion projects in the USA, the UK, France and elsewhere as well  
as special agreements like the one between Russia and Italy on alternative costly fusion projects. Obviously such projects raise doubts about the scientific 
foundations of the ITER project.
\item The distribution of the knowledge that the ITER project has absolutely nothing to do with commercial 
energy production and that, even if realized according to plans, not even some of the most basic tests 
required for a future even bigger fusion project can be performed. 
Among these tests are the tritium breeding problem and 
the need to develop a neutron radiation resisting material which at the same time can survive the always occurring plasma 
eruptions. These and other fundamental problems have been 
explained in detail in the article ``Fusion Illusions"\cite{fusionillusions}. As explained, enough knowledge about  
the imagined tritium breeding process, required for a future commercial power plant, has been accumulated to understand that 
nobody within the fusion community has even the slightest idea on how this problem can be solved. 
One can be sure that once this still-well-hidden problem becomes common knowledge, the enthusiasm 
from the entire scientific community will disappear quickly. 
\item The recent statement from French physics Nobel laureat G. Charpak 
and other scientists asking for an end of the ITER project and for the transfer of the money to 
more down-to-earth nuclear fission projects\cite{charpaketal}.
\end{itemize} 

The year 2010 might not be the termination year of the ITER project, but perhaps it can be defined as the year when the ``false dawn" of nuclear fusion 
was realized. We can thus safely predict that the belief in commercial nuclear fusion on our planet will end once the 
younger generation of scientists sees that plasma fusion research is 
a dead end career path and turns its talents to other research projects.

% stop 29.9.2010

\section{Energy from nuclear fission: the years 2010-2015(20): \\
Some emerging patterns from the past five to ten years}

In comparison with other industrial areas, the entire nuclear energy chain is known for its long planning and construction times. 
On average it takes at least 5 years to complete a 1 GWe nuclear power plant, uranium mines and other needed facilities.  Depending somewhat on the country, similar
times are needed prior to this for the planning phase of nuclear power plants.  Programs for new types of reactors, like the Generation IV initiative, the search for a final nuclear waste 
storage and the ongoing struggle with the ITER plasma physics project point to timescales of at least 20-30 years.  
It follows from these long construction schedules that the maximum possible contribution of nuclear energy during the next 5-10 year was essentially fixed during the past 5-10 years.

In short, if every one of the more than 60 reactors currently under construction can be realized according to schedule, the nuclear power capacity by 2015 might reach at most about 425 GWe, roughly a 15\% increase from todays capacity of 374 GWe. 
The situation for the years 2016-2020 and thereafter becomes less clear but the different editions of the Red Book provide a summary of the 
plans from different countries now up to the year 2035\cite{RB09}. The plans for some countries with larger changes are summarized in Table 1.
The accuracy of those plans can so far only be compared to this year and, like in past years, the reality shows that even the just published 
2009 edition of the Red Book is between 5-17 GWe too high. The reason to publish totally impossible numbers for the year 2010  must be that some country correspondents provide knowingly false estimates and the editors of the Red Book, even if they know better, can not change these numbers.  When looking 10 to 15 years forward, it is inevitable that  uncertainties become larger. 
It might be interesting to compare the 1997 Red Book prediction for the world nuclear capacity in 2010 and 2015 with the reality.
The prediction for 2010 was a world wide nuclear capacity of 427-461 GWe and between 394-501 GWe by 2015.  
We leave it to the reader to judge the reliability of the nuclear capacity predictions for 2015, 2020 and beyond. 

{\small
\begin{table}[h]
\vspace{0.3cm}
\begin{center}
\begin{tabular}{|c|c|c|c|}
\hline
source                     & power    [GWe] 2010                  &  power [GWe]  2015                      & power [GWe]  2020      \\
\hline
PRIS (Aug 2010)                               &    374.6                   &                                          &                           \\
\hline
RB97 world total                      &  427-461                &   394-501                            &                          \\
RB07 world total                      &  377-392                &   410-456                            &   450-516         \\
RB09 world total                      &  381-393                &   409-442                           &   450-535         \\
required uranium/year                  &  68 ktons  (WNA 2010)  &  72-77 ktons          &  79-94 ktons    \\
\hline
RB07 OECD total                      &  304-309                &   310-326                           &   315-341         \\
RB09 OECD total                      &  312-315                &   314-323                           &   320-353         \\
\hline
RB09 (WNA Aug 2010 )   & power    [GWe] 2010                  &  power [GWe]  2015                      & power [GWe]  2020      \\
\hline
China  (9.6 GWe )                  & 13-20                                            & 25-35          &  40-58   \\
India  (4.2 GWe)                      &  5.2-6.2                                        &  8.1-13.2      & 11.6-21.3  \\
Japan (47.3 GWe)                     & 48.0                                          & 48.3-52.2        & 53.7-64.0  \\
South-Korea (17.7 GWe)                 & 18.7                                  & 25.9               & 31.5   \\
Russia (23.0 GWe)                    & 22.7-23.4                           & 28.1-30.9            & 34.7-41.2        \\
\hline
Germany  (20.4 GWe)            & 20.5                                        & 12.1-13.4         &   3.5    \\
UK (11.3 GWe)                        & 10.5                                          & 7.2                  & 4.4-5.8        \\
\hline
sum  (133 GWe)                         &  139-147.3                           &   155-178        &  179-225       \\               
other (240.2 GWe)                      &  242-248                           &   254-264        &  271-310       \\               
\hline
\end{tabular}\vspace{0.1cm}
\caption{Nuclear power perspectives up to 2020 for different countries and the world as given 
by the Red Book 2009 (RB09) and 2007 (RB07)\cite{RB09}. The 2010 numbers from PRIS \cite{PRIS} and from the WNA \cite{WNAnewreactors}
demonstrate that even the Red Book 2009 numbers for 2010 are between 5-17 GWe too high. The natural uranium requirements have been simply calculated 
from using the RB09 power estimates for 2015 and 2020 and with 175 tons per 1 GWe nuclear power plant. The additional amount required for the first load of new reactors has been ignored.
Mining needs to produce the estimated amount minus the amount coming from secondary resource, currently estimated to decrease from roughly 20000 tons to about 10000 tons after 
2013.}
\end{center}
\end{table}
}

% \normalsize

As the aging of the world reactor fleet continues, one can expect that their performance will become less and less reliable as already documented 
with the evolution of average production performance, which has decreased since 2006 on average by about 1\% per year.
In addition some reactors will also be terminated during the coming years and, 
taking the pattern from the past 5-10 years, one could expect an average yearly reactor termination, with a power loss of 1-1.5 GWe. 

In any case, the existing and future nuclear reactors need to be fueled with about 170 tons/GWe of natural uranium equivalent per year. 
Following the latest estimates about the decreasing availability of secondary uranium resources to about 10000 tons after 2013, 
the world uranium mines need to extract a total of about 55000 tons of 
natural uranium to fuel the existing capacity of 374 GWe. If the additional power plants of 50 GWe can be realized by 2015 
the mines need to provide an additional amount of at about 10000 tons, ignoring the additional amount for the first fuel load.

In the following subsections these overall prospects for nuclear power and the corresponding uranium fuel needs 
are presented for different countries and regions starting with the announcements about uranium mines.

\subsection{Announcements for new uranium mines} 

According to the 2009 edition of the Red Book, new large uranium mines with a production capacity of more than 2000 tons are
supposed to start during the next 5 years in Kazakhstan (plus 10000 tons), Namibia (plus 9200 tons), Niger (plus 5000 tons), Canada (plus 6900 tons) and Jordan (plus 2000 tons). 
The total production capacity of the 9 larger mines adds up to 33000 tons by 2015. In addition, many smaller mines with capacities between a few hundred to 2000 tons 
will open in several countries, adding perhaps another 16000 tons of capacity by 2015. 
 
Assuming that the existing capacity, given for 2010 as 70-75000 tons,  
will not be subject to significant declines, the authors of the Red Book 2009 estimated the total maximum uranium mining capacity for 2015 to be between 96000 tons and 122000 tons.
The predicted spread indicates the large uncertainties about the new projects and the date when full capacity can be reached. 
When looking further forward, an important prediction about the peaking of the mining capacity is made by the Authors of the 2009 Red Book. It is stated that 
the worldwide uranium mining capacity is expected to reach a maximum between 98000-141000 tons
around the year 2020, followed by decline to 80000-129000 tons (2025), 75000-119000 tons (2030) and 68000-109000 tons (2035). 
As if these numbers would not be surprising enough, the ones given for Kazakhstan can perhaps be called alarming. 
A peak production of 28000 tons is expected already around the year 2015, followed by a decline to 24000 (2020), 14000 tons (2025), 12000 tons (2030) and to 5000-6000 tons by 2035.  

{\small
\begin{table}[htb]
\vspace{0.3cm}
\begin{center}
\begin{tabular}{|c|c|c|c|c|}
\hline
Red Book                     &  prod. capacity                     & prod. capacity         & prod. capacity       & prod. capacity \\
2009                             & 2010 [1000 tons]                 &  2015  [1000 tons]   &  2020  [1000 tons]  &  2025  [1000 tons]\\
\hline
Australia                       & 9.7                           &    10.1-16.6                      &    10.1-24.2        &         10.1-27.9                                     \\
Canada                        & 16.4                          &    17.7                             &  17.7-19.0          &          17.7-19.0                                    \\
Kazakhstan                  &  18.0                           &     28.0                    &  24.0                           &           14.0                     \\
Namibia                       & 5.0-6.5                        &      6.0-15.0               & 8.0-19.0                &           6.0-14.0                                        \\
Niger                           &  4.0                         &         9.5-11.0                    & 9.5-10.5            &           5.0-9.5                                         \\
Russia                         & 3.5                           &         5.2-5.9               & 7.6-12.0                 &           7.6-13.8                     \\
South Africa                  & 4.9                 &          4.9-6.3                       & 4.9-6.3                    &           4.9-6.3                            \\
USA                            &   2.9-4.6           &          3.4-6.1               & 3.8-6.6                           &          3.7-6.5           \\
\hline
World                           & 70.2-75.4                            &  96.1-121.8             &   98.3-140.6       &         79.7-129.3         \\ 
World (new cap.)         &      4                                     &  26-52                     &    28-70.4           &          29.5-59.1          \\ 
\hline
Scenario A             &   53-56.5             &  72-91.3                  &    73.7-105.5      & 59.8 -97.0                                     \\
Scenario B             &   52             &  63-76                  &    64-85      & 65-80                                                               \\
\hline
World 2009        &   51               &                    &          &                                                                \\
World 2010(?)        &   55               &                    &          &                                                                \\
\hline
Kazakhstan 2009    &   14               &                    &          &                                                                \\
Kazakhstan 2010(?)  &   18               &                    &          &                                                                \\
\hline
all other 2009        &   37               &                    &          &                                                                \\
all other 2010(?)        &   37               &                    &          &                                                                \\
\hline
\end{tabular}\vspace{0.1cm}
\caption{Expected uranium production capacity given in 1000 tons from the Red Book for world and for different countries and for 
the years 2010 and 2015\cite{RB09}. The expected world wide capacity increase between 2007 
and 2010 and from 2010 to 2015 are obtained from the evolution of the total capacity.
Scenario A and B are a rough forecasts for the maximal uranium mining for the years 2010 and 2015 and based on the past capacity and real mining relation.
For Scenario A it is assumed that the mining performance will be 75\% of the future capacity expected according to the Red Book.  
For Scenario B we assume that the existing mines in 2009 will continue an average annual production of 50000 tons and  
that only 50\% of the capacity forecast can become operational in time. }
\end{center}
\end{table}
}

The authors of the Red Book acknowledge that the uranium mining capacity numbers are very different from the much lower real mining production. 
In order to quantify this difference, one can confront the evolution of the mining capacity numbers from the different editions of the Red Book with 
the almost known mining result for the year 2010. 
Accordingly, past capacity predictions for the year 2010, were 69000-83000 tons (Red Book 2005) and 81000-96000 tons (Red Book 2007). 
Taking the latest and naturally more accurate Red Book 2009 estimate for 2010 of 70000-75000 tons, which is 11000-21000 tons smaller than the prediction from 2007, it follows that roughly 50\% of the ``new" projects have ``evaporated" during the last two years.
 
In addition and according to the latest estimates, uranium mining in 2010 is expected to reach about 55000 tons.
The real mining is thus 15000-20000 tons, or 20-30\%, smaller than the mining capacity figure given in the just published 2009 edition of the Red Book.

In chapter I of ``The future of nuclear energy" from 2009\cite{oildrumart}, we had presented two scenarios, A and B, to guess the future evolution from past trends. 
According to scenario A, which assumes that about 75\% of the Red Book 07 capacity can be realized, a 2010 estimate of 
60500-65000 tons was obtained. For scenario B it was assumed that the existing mines continue their annual production of 40000 tons (roughly the 2005 result)
and that only 50\% of the new mines can be realized in time and a forecast for 2010 of 53000-55000 tons was obtained.
Scenario B is thus in agreement with the 2009 and most likely also with the 2010 result.  

Following the same procedure to estimate the uranium production for the year 2015  
and with the 2007 Red Book capacity estimates, the predictions were 72000-88000 tons (scenario A) and 61500-70000 tons (scenario B).
If one uses the 2009 Red Book capacity estimates for the year 2015 and takes the 2009 mining result of 50000 tons as an updated 
baseline, the expected 2015 mining would be 72000-91000 tons (Scenario A) or 68000-78000 tons (Scenario B). 

In addition to these two scenarios my own 2009 guesstimates, 
assumed that uranium mining will be strongly affected by the financial crisis and by the increasing troubles to exploit lower and lower grade. 
My scenario was wrong for Kazakhstan during 2009 and 
2010\footnote{The almost too perfect matching of the yearly mining results with the plans from the past years 
reminds me of some historical 5-year plans in this area. Therefor I remain rather sceptic about the current and future claims from uranium mining in Kazakhstan.}.  
On the other hand, my last year guesstimate was in rough agreement with the mining results from all other countries in 2009 and presumably also with the one in 2010.

Therefore, and despite the fact that my 2009 uranium mining estimate was 7000 tons too low, my 2010 updated guesstimate for the next years is that  
primary uranium production from all countries, other than from Kazakhstan, will not increase significantly above 40000 tons during the coming five years. 
I thus continue to predict problems with uranium supply and assume that shortages after 2013 can only be avoided if new ``miracles" (at least for me)   
will happen. Examples of such ``miracles"  are that uranium mining in Kazakhstan can really be further increased to the claimed 24000 tons by 2015
or that a new deal concerning the military uranium reserves between Russia and the USA will be made in time. 

Some quantitative Red Book numbers for the total uranium production capacity and for a few large uranium producing countries for the years 2010, 2015, 2020 and 2025 are summarized in Table 2.

\subsection{Can new reactors compensate the effects from overall aging?}

The IAEA PRIS reactor data base \cite{PRIS} documents the performance of the individual nuclear reactors 
and gives some overall nuclear energy figures. Between the year 2004
and 2009 one finds that 16 new reactors have started operation.
During the same period, 19 older and mostly smaller reactors have been terminated after an operation of at most 40 years. 
Taking the abnormal long construction times of some Chernobyl type reactors and of the one in  Iran out of the equation, the construction of a modern nuclear power plant takes 
5-8 years from initiation to the first grid connection and about one more year to reach full performance.
When looking at the individual performance 
of the new reactors, one finds that it took a few months to about a year for half of the newly grid connected reactors to reach the nominal capacity with an average above 85\% operation time. 
For the other half it seems that after the official grid connection date it takes a few years to reach nominal power and this with an operation time between 50-80\% only.  
Perhaps somewhat surprisingly, the highest performance figures for new reactors were obtained by the three new large reactors in China and the one in Rumania. 

\begin{itemize}
\item Following the pattern from the past 6 years one might expect that the construction times for 60 reactors, currently under construction,
especially for the ones in China, Japan and South Korea will be about 5-7 years, roughly 20\% longer than in the original schedule,
assuming that no additional financial crisis or other types of ``problems" will happen.  
\item New reactors are operated on average for about two years with a 60\% capacity 
before they reach current world average reactor availability factors about 80-85\%. 
In other words, average world wide reactors are running at about 100\% capacity during about 80\% of a year. 
Taking this number,  a 1 GWe nominal power reactor might produce 7 TWhe electric energy per year while newly grid connected 
reactors can be expected to produce about 6 TWhe during its first full operational year.
\end{itemize}

The Tomari 3 reactor connected to the grid in Japan (20 March 2009) 
produced 2.43 TWhe during its first 9 months of operation~\cite{PRIS}. According to the above guessed early performance figures, a production of about 5 TWhe
can be expected in 2010 from this reactor. Adding the other newly connected reactors in Russia, India and Korea one could 
assume that the new reactors together add another 5-8 TWhe during 2010. 
However, the 2009 running of the now terminated Ignalia II reactor in Lithuania produced 10 TWhe~\cite{PRIS} and we can thus expect 
at best a break-even between new and terminated reactors since 2009.     
 
Concerning the scheduled startups of about 10 new reactors per year during the coming 5 years, we 
predict that these aggressive timescales can not be realized. This view is backed up by  some delays which have already been announced by  
several countries~\cite{delays}.

Thus, we guesstimate the future nuclear capacity as follows:
\begin{itemize}
\item Only half of the promised startups per year will happen according to the official plans   
and up to 2015. Including some upgraded capacity plans, this would still correspond to a yearly worldwide capacity growth of about 5 GWe.
\item Without well defined dates for reactor terminations we make the working hypothesis that  
shutdowns will continue with an average capacity loss of about 2 GWe, following the pattern from the past 5-10 years.
\end{itemize}
It follows that a net capacity increase of 3 GWe per year can be expected,
reaching perhaps a total capacity of 390 GWe at the end of 2015.

In order to guess the real electric energy production from the world nuclear fission power plants during the next couple 
of years one needs to take the average energy ``availability factor" of the aging reactors into account.
During the year 2009 this availability factor was 79.4\%, the lowest in the past decade and on average the performance of the 
reactors has declined by about 0.5-1\% per year since the year 2004. 
As the aging of the nuclear power plants will continue during the next 5 years, several additional unplanned ``repair" outages can be expected. 
We thus assume that the overall availability factor of the existing reactors will continue to decrease by perhaps 0.5\% per year. 

Taking all these factors and the large uncertainties into account, we guess that the electric energy production from the worldwide nuclear power plants will stagnate 
around the numbers achieved during the past decade, somewhere between 2500-2650 TWhe per year.  

Of course, a potential uranium shortage after the year 2013, following the end of the 10000 tons of Russian military uranium delivery to the USA, 
and other unexpected smaller or larger ``incidents" affecting the entire nuclear fuel chain can only result into a stronger decrease to numbers as low   
as 2350 TWhe for the year 2015.

\subsection{Regional perspectives for nuclear energy}

The commercial use of nuclear fission energy started some 50 years ago and today 441 nuclear power plants 
with a total electric capacity of 374.6 GWe are considered as operational in 30 countries. Currently 344 reactors with a capacity of about 315 GWe, about 84\% of the total, 
are operated in the relatively rich and highly industrialized OECD countries in Western Europe, North-America, Japan and the Republic of Korea.
Some nuclear energy saturation effects and even declines are observed and more are expected for the OECD regions which currently depend on a relatively important amount of nuclear produced electric energy. 

The picture is drastically different when one looks at the 60 reactors currently under construction. 
Most of them are found in the larger countries with a predicted continued fast economic growth. 
It follows that the observed stagnation of nuclear power in the OECD block is confronted with very ambitious plans for new nuclear reactors in 
China, Russia and India. These plans seem to have some parallels with similar past aggressive nuclear growth scenarios in the OECD countries following the 
political oil crisis during 1973. The fact that not even 50\% of these aggressive nuclear growth scenarios were realized in the past is a warning for those who believe today's predictions about strong nuclear growth scenarios. 
Some more details about the plans for future nuclear power plants are summarized in Table 1 above. 

An overview about the actual situation and the short term perspectives in regions with common economic interests are presented in the following subsections. 

\subsubsection{Europe (OECD)}

About 150 reactors with a capacity of 135 GWe are currently in operation. 
These reactors, currently well integrated in the european grid system, provided 830 TWhe in 2009, about 25\% of the total produced electric energy.
This number is significantly smaller than during the year 2006 when 930 TWhe, corresponding to a fraction of 27\%, were produced. 
 
Two EPR reactors, from the French state company AREVA, are under construction in Finland and in France. A few more countries are reported as being in a serious planning phase for replacement of their 30-40 year old reactors. 

Significant uranium mining in Western Europe ceased during the last decade of the 20th century and without significant civilian or military secondary reserves,
the european countries are essentially 100\% dependent on uranium imports.

Given that a large fraction of Europe's 150 reactors will soon reach the end of the originally licensed lifespan, but given also  
that the import dependence on oil, gas and even coal looks more and more problematic for coming decades, there is considerable uncertainty 
and even confusion, regarding the future of nuclear energy in Europe. The latest decisions in Germany about delaying the ``nuclear phase out"
are still somewhat ambiguous and the ongoing heated debate is far from declining\cite{germany}.

However, perhaps the most ``official" numbers about the future nuclear power capacity can be estimated from the EURATOM uranium supply agency, 
which estimates in their 2009 annual report\cite{Euratom}, that the uranium equivalent reactor needs will decline from 21800 tons in 2010 to 20000 tons between 2015-2020 and further to 
18000 tons by 2025. This decline corresponds to a 5-15\% reduction of nuclear power during the next 5 to 15 years, an average yearly  
reduction of about 1.3 GWe. 
 
Regarding nuclear waste\cite{wasteprojects}, Finland has begun what is probably the world's most advanced final deep underground storage project: a deep underground cavern which might start operation by the year 2020.  Decision makers in other European countries appear to be counting on future generations 
or present-day miracles to solve the problem of nuclear waste. For example some projects, like the German Gorleben salt cavern waste storage project, 
are plagued with ongoing long term safety concerns\cite{Gorleben}. 
Assuming that no miracles are seen during the next 30 years, the most likely final solution --or more accurately, final nonsolution\footnote{Of course the simplest and cheapest option for the 
current generation of nuclear electric energy users.} -- will be that the nuclear waste will remain for some decades next to the nuclear power plant (and their future ruins) and will at some point  
begin to contaminate the groundwater nearby and probably also the ``downstream" cities and countries. 
 
\subsubsection{North-America (OECD)}
There are 124 nuclear reactors operating in this region and one that will be soon operating:
104 reactors, plus 1 under construction, in the USA, 18 in Canada and 2 in Mexico. About 20\% of all electric energy in this block is nuclear.
In 2009, electric energy from nuclear power amounted to 895 TWhe, about 1\% lower than in the record years of 2007 and 2008.
The preliminary data from the first few months of 2010 indicate a further decline of 2\% during the first eight months compared to 2009.  

While there is talk of increasing capacity by perhaps 5\% in this block over the coming decade, the reality is that few reactors 
are presently in a serious planning phase. 

The natural uranium equivalent requirements from these three countries for 2010 are 19538 tons for the USA, 1675 tons for Canada and 253 tons for Mexico.

Concerning uranium mining, Canada accounted for about 20\% of total world production in 2009, about 10173 tons, a large fraction of which went 
to the USA and Western Europe.
Especially interesting is the uranium supply situation in the USA. Currently only 8\% of their uranium needs, 1453 tons in 2009, are coming from their own mines. 
This is especially remarkable (1) as uranium mining capacity used to be the largest in the world, reaching an annual record production of almost 17000 tons some 30 years ago 
and (2) because energy independence has been one of the stated goal of essentially all American governments for decades now. 

This dependence is even more interesting as 50\% of the current USA uranium needs are satisfied   
with a yearly import of roughly 10000 tons from Russia. As this yearly delivery will stop at the end of 2013,   
the US government faces some very tough decisions regarding civilian use of its roughly 200000 tons of military reserves in coming years.

The situation regarding the final deposit of the nuclear waste is, similar to that in Europe.  Decision makers are essentially relying on present-day miracles or future generations. 
This view was strengthened after the advanced Nevada Yucca Mountain deposit project was abandoned in spring of 2009 by the Obama administration,
a decision made after about 13 billion dollars, of a total estimated cost of 96 billion dollars,  had already been spent \cite{YuccaMountain}.
 
\subsubsection{Japan and South-Korea (OECD Pacific)}

The number of commercially operating nuclear power plants in Japan and South-Korea is 54 and 21 respectively. 

In 2009 the 54 Japanese reactors produced about 27\% of the country's total electric energy. Japan is still recovering from the forced shutdown 
of several reactors after a strong earthquake in 2007: 2006 was a record production of 301 TWhe; 2007 dropped to 268 TWhe; 2008 dropped further
to only  246 TWhe; 2009 was similar to 2007 at 267 TWhe; and a further recovery, about 7.3\% higher than in 2009 has been reported for the first 
eight months of 2010 \cite{IEAdata}.

The Monju prototype fast breeder reactor, currently the only one in the world, 
has just been restarted\footnote{This restarting has been interrupted by an incident in August 2010, when a heavy element was dropped into the reactor.}
for experimentation after a long shutdown.
Two reactors (not the Monju type) are now under construction and a few more are in a planning phase. 

South-Korea currently operates 21 reactors, 5 more are under construction and 6 in a more serious planning phase. 
These reactors produced 141 TWhe in 2009, slightly lower than during 2008 when 144 TWhe were produced.
The results from the first eight months of 2010 are 0.3\% lower than in 2009. 

Neither Japan nor South-Korea has ever had any significant uranium mining and both are thus 100\% dependent on imported uranium which, for 2010,
will amount to 8000 tons and 3800 tons respectively. 
Although a large fraction of their uranium currently comes from Australia, it will soon also come from Kazakhstan and Russia. The growing competition with China, and potentially also 
India, for uranium imports is certain to become problematic during the coming decade. 

The problem of nuclear waste is without solution in both countries and is almost certain to be handed over to future generations, as is the case almost 
everywhere else.

\subsubsection{Russia and Ukraine}

Nuclear energy production in these countries has an unfortunate legacy from the days of the Soviet Union and the sort of 
reactors that led to the Chernobyl disaster in 1986. 
Currently 31 and 15 reactors with a capacity of 21.7 GWe and 13.2 GWe are operated in Russia and in the Ukraine respectively.
 
Ukraine's reactors produced 78 TWhe in 2009 --roughly 10\% less than in 2007,  but still about 48\% of the country's electric energy. Russia on the other hand achieved record production in 2009 --163 TWhe, about 16\% of the country's produced electric energy --thanks to improved reactor performance.

Both countries are aggressively planning to at least double the nuclear capacity during the next 10 years and many more reactors are considered to be in a planning and proposal phase.
However, several construction delays have already been announced and the latest plans indicate that a real capacity increase can start only after the year 2015. For example according to the WNA document, although the situation in Russia is rather unclear, it looks like at best 
only about 4 GWe might be added by 2015\cite{WNArussia}.
 
The situation in the Ukraine is that plans for the initiation of at most 6 reactors up to 2015 exist, resulting in an earliest possible start only 6 years later.  
It is thus unlikely that a new reactor can be grid connected before 2017\cite{WNAukraine}.

Uranium requirements for 2010 are 4135 tons in Russia and 2031 tons in the Ukraine.  Uranium mining during 2009 corresponded to 3564 tons in Russia and 840 tons in the Ukraine.
Both countries share a common ``friendly" history --and Russia has especially 
strong ties to-- Kazakhstan, the rising star of uranium production, and both Russia and the Ukraine are also on very good terms 
with Uzbekistan which produced a non negligible amount of 2500 tons of uranium in 2009.
A large new uranium mining area in Siberia with a yearly capacity of 5000 tons is currently planned to start in time to satisfy Russia's needs beginning in 2015.

In addition, huge secondary reserves of about 300 000 tons are attributed to the former Soviet Union states dominated of course by Russia. 
Essentially all of it can be assumed to be currently controlled by the military and thus by the Russian government. 
The importance of these huge reserves, combined with the strong Russian ties to uranium mining in Kazakhstan, will grow during the coming years.
A possible result might be that all OECD countries without access to uranium resources, will be forced into an even stronger Russian dependence.

The storage of nuclear waste in these countries is rarely discussed, but some regular appearing news about the simple dumping of military 
nuclear submarines into the sea and other doubtful practices is certainly worrying\cite{russiawaste}. 

\subsubsection{China and India}

The two most populous countries on the planet, China and India, have extremely ambitious plans for an increase of nuclear energy production. 

China with a population of 1.34 billion people has a yearly per capita production of about 2500 kWhe.
This number has roughly doubled in parallel with the strong economic growth since the beginning 
of this century, but is still a factor of three below the one in richer European countries.

Nuclear power currently plays an almost negligible role: the 12 nuclear power plants in China with a capacity of 9.6 GWe produced 65.7 TWhe in 2009, about 1.9\% of the total electric energy. 
Following its economic success and the accumulation of up to about 1000 billion dollars, China is now in a position to simply buy -- in the ``free market" --
many high tech items like nuclear power plants, large amounts of uranium and many other vital resources.

Since China now faces enormous environmental problems created by their extensive use of dirty coal, 
the option of increasing nuclear power production appears to be logical.
In addition the operation of nuclear energy is still considered 
as one of the most obvious signs for a successful transition to a modern society.  
The result is thus that 26 nuclear power plants are currently under construction in China.  
If their construction proceeds according to plans about 2-3 GWe of new capacity will be added per year up to 2012 and another 6-7 GWe per year  
from 2013 to 2015. Plans exist to continue this growth with 33 more reactors by 2020 and another 120 reactors for the years 2020 to 2030. 
According to the latest plans, the nuclear power capacity should thus reach about 35 GWe by 2015, 80 GWe by 2020, 200 GWe by 2030,
roughly twice the number in the USA today, and perhaps even 400 GWe by 2050\cite{WNAchina} 

These plans require corresponding large amounts  
of uranium, starting with a  demand of 2875 tons in 2010 to perhaps 6000 tons by 2015, 14000 tons by 2020 and 30000 tons by 2030. However, the 2009 uranium mining in China 
produced only 750 tons and without any significant uranium deposits of its own, China has been aggressively investing in existing 
and planned uranium mines all around the world. 

The situation in India shows many similarities with China, but this with a delay of roughly a decade. India's 19 nuclear power plants have currently a capacity 
of 4.2 GWe and produced 14.8 TWhe in 2009, about 2.2\% of the country's electric energy. The per capita electric energy consumption 
is still roughly a factor of three smaller than in China.  
The relative economic success during the past few years resulted in the increasing strength of the country and corresponding ambitious plans to increase nuclear power 
during the next 10-15 years. 

Currently 6 reactors with a capacity of 3.8 GWe are under construction, 20 are in a more serious planning phase and another 
40 are in a proposal phase. If all these reactors would be realized within the next 15 years, the total nuclear power capacity of India would reach about 70 GWe, thus more than in France today\cite{WNAindia}.

The current poor performance of the existing reactors is partially explained by uranium shortages due to difficulties obtaining uranium imports --
difficulties that stem from the fact that India is a nuclear weapon state 
outside of the NPT treaty. Such difficulties have slowed but not stopped imports: several countries including even USA under the Bush administration,
have made it possible for India to buy uranium on the international market, 
even though this is an apparent violation of NPT treaty obligations.  
Like China, despite its large territory India manages to produce only a small amount of uranium itself Ð 290 tons of the needed 908 tons in 2010 for example.

Wether and how the future uranium needs of China and India can be satisfied -- even if almost all future production went to these countries -- 
remains to be seen. If these needs can not be satisfied, perhaps their dreams of becoming rich industrialized countries, 
thanks to billions of dollars spent on nuclear power plants,  will end in nightmares with 
unfinished or unfueled plants lying in ruins. 
 
The situation of nuclear waste in China and India matches those in Russia and the Ukraine.
It is rarely discussed, but the news about many other huge environmental problems in these countries indicate that the coming additional nuclear waste problems might lead to new disasters 
for the concerned populations. With respect to the recent flooding catastrophes one could guess that this and other nightmares are just another flood or earthquake away. 
%  stop 2.9.2010 
\section{The 2009 Red Book uranium resource data, more unreliable than ever?} 

The latest 2009 edition of the Red Book uranium resource data was published on July 20, 2010\cite{pressconfRB09}. 
In this book,  updated every two years,  the IAEA (International Atomic Energy Agency) from the United Nations 
and the NEA (Nuclear Energy Agency) from the OECD countries have, for more than 40 years \cite{RB40years}, presented their 
collective knowledge about uranium resources and their use for civilian nuclear energy. The data 
are widely perceived as assuring the viability of future nuclear energy projects. 
  
Unfortunately, the data presented in the new Red Book edition
do not measure up to the pretended standards. They are in fact of such poor quality that any perceived assurances are likely to lead to nothing but disappointment regarding the uranium supplies that will be needed in future decades to operate all of the expensive new nuclear power plants that various countries are planning to build, and all of the existing plants still in operation.

What follows will address the key findings only.  A more detailed analysis of the new Red Book's uranium resource data will follow at a later date.

As in past years, a fairly euphoric press release introduced the latest edition (italics added and bolding in the original): 
 
{\it ``The uranium resources presented in this edition, reflecting the situation as of 1 January 2009, show that total identified resources amounted to 6 306 300 tU, an increase of about 15\% compared to 2007, including those reported in the high-cost category ($<$ USD 260/kgU or $<$ USD 100/lbU3O8), 
reintroduced for the first time since the 1980s...."}  

--Followed by:

{\it ``Although total identified resources have increased overall, there has been a significant reduction in lower-cost resources owing to increased mining costs. {\bf At 2008 rates of consumption, total identified resources are sufficient for over 100 years of supply}...."} 

--And: 

{\it {\bf ``Even in the high-growth scenario to 2035, less than half~\footnote{A remarkable statement that during the next 25 years close to half of the optimistically claimed resources are already gone. 
Not very assuring for future buyers of reactors, with a 60 year lifetime, that their expensive marvel might receive fuel only up to around 2060.}
of the identified resources described in this edition would be consumed. The challenge remains to develop mines in a timely and environmentally sustainable fashion as uranium demand increases.} A strong market will be required for these resources to be developed within the time frame required to meet future uranium demand."} 

The cited parts of the press declaration and especially their bolded statements can be considered as an invitation to look inside the report to find their justifications.  

The following distinctions are made in the Red Book regarding uranium resources: known deposits where there is some reason for confidence regarding 
mineable tons and grade are called Reasonably Assured Resources (RAR) while not yet discovered but believed to 
exist deposits are called Inferred Resources (IR); these two together are viewed as total conventional resources. 

In addition, there is an iffier category called Undiscovered Prognosticated Resources (UPR) and an even iffier category called Undiscovered Speculative Resources (USR); both of these are so speculative that serious plans for future use should not count on them.  As in past editions of the Red Book, these various categories are further divided according to estimated mining cost ranges.  The data are presented with an impossible to know pseudo accuracy 
and without an uncertainty range. It follows that most of the presented figures are not based on geological evidence but more on 
bureaucratic accounting schemes.   

\subsection{A 15\% increase of the conventional uranium resources?}

The reported overall increase of conventional uranium resources, the sum of the RAR and IR categories, from 2005 to 2007 and to 2009 corresponds to an overall increase of 20\% and 15\% respectively
during subsequent 2 year periods.  It is interesting to understand where and in which price category the supposed increase has occurred. 

Tables 3-6 summarize the world total resource estimates for the different categories 
and their evolution as given in the last 5 Red Book editions   
from 2001, 2003, 2005, 2007 and 2009\cite{RB09, RB40years}.
  
In order to simplify the discussion, the numbers are recalculated such that the uranium amounts for a given cost interval can be compared. Table 3 shows the evolution of the conventional resources 
since 2001. The ``huge" increase during the last two years comes only from the 
newly introduced high cost category. As this new cost category has been introduced only in order to take increases in extraction costs into account, 
the origin of the additional 0.9 million tons is mysterious.

{\small
\begin{table}[htb]
\vspace{0.3cm}
\begin{center}
\begin{tabular}{|c|c|c|c|}
\hline
Red Book year      & RAR  [tons]                     & IR  [tons]             & conventional resources [tons] \\
                             & $<$ 130 dollars/kg          & $<$ 130 dollars/kg             & $<$ 130 dollars/kg   \\
\hline
2001                    & 2 853 000               &   1 080 000                           & 3 933 000  \\
2003                    & 3 169 238               &   1 419 450                           & 4 588 688  \\
2005                    & 3 296 689               &   1 446 164                           & 4 742 353  \\
2007                    & 3 338 300               &   2 130 600                           & 5 468 800  \\
2009                    & 3 524 900               &   1 879 100                           & 5 404 000  \\
\hline
Red Book year      & RAR  [tons]                     & IR  [tons]                     & conventional resources [tons] \\
                             & 130-260 dollars/kg          & 130-260 dollars/kg      & 130-260 dollars/kg   \\

2009*                   &   479 600               &     422 700                             &    902 300  \\
\hline
\end{tabular}\vspace{0.1cm}
\caption{The evolution of the conventional uranium resources split into the 
reasonably assured resource (RAR) and the inferred resource (IR) categories from the latest five 
Red Book editions.  For the 2009 edition a new category, from 130-260 dollars/kg, has been introduced. 
The number of tons for this category are listed under 2009*.  Thanks to the addition of this new category, total conventional resources for all cost levels are now said to be 6 306 300 tons.
Especially remarkable in the latest edition is the increase in the RAR number by almost 200 000 tons and the decrease 
in the IR number by a similar figure, while the overall RAR + IR number for the same cost category remains almost unchanged. 
}
\end{center}
\end{table}
}

Table 4 and 5 show the corresponding evolutions for the RAR and IR categories, 
split according to the estimated extraction cost range. As in the past, there is no way of knowing whether the supposedly precise RAR numbers are adjusted for the yearly extraction of roughly 40 000 tons.  One can only hope that they are.

{\small
\begin{table}[h]
\vspace{0.3cm}
\begin{center}
\begin{tabular}{|c|c|c|c|}
\hline
Red Book year      & RAR  [tons]                                    & RAR  [tons]                           & RAR [tons] \\
                             & $<$ 40 dollars/kg          & 40-80 dollars/kg             & 80-130 dollars/kg   \\
\hline
2001                    & 1 534 100               &     556 650                             &    589 770  \\
2003                    & 1 730 495               &     575 197                             &    661 941  \\
2005                    & 1 947 383               &     695 960                             &    653 346  \\
2007                    & 1 766 400               &     831 600                             &    740 300  \\
2009                    &    569 900               &  1 946 000                             & 1 000 900  \\

\hline
\end{tabular}\vspace{0.1cm}
\caption{The evolution of the reasonably assured resource (RAR) category from the latest five 
Red Book editions.  Especially remarkable is the rather sudden and dramatic decrease 
in the lowest best known cost category during the last two years. 
After the regular large increases and a sudden decrease by about 180 000 tons for the 2007 edition, one finds that 2/3 of the uranium number from this cost category 
has been moved to the more expensive ones and an additional (out of the hat?) amount of 479 600 tons is 
assigned to the new high cost category of 130-260 dollars/kg. As this period coincides with the period when the oil price increased by a large amount
one might guess that the mining costs are strongly correlated with the price of oil. }
\end{center}
\end{table}
}

Furthermore, the RAR numbers, claimed to be known with great precision, 
appear to fluctuate by large amounts. Between 2001 and 2005 the best understood (and most interesting for mining) 
category of $<$ 40 dollars/kg category increases by roughly 200 000 tons every two years, while the more expensive ones remained essentially unchanged.
However, starting in 2007, an almost dramatic transition from the 
best understood category to the less well understood higher cost categories is reported.
Such changes could perhaps indicate that the authors of the Red Book acknowledge that  their understanding about uranium resources is far below the one  
which was claimed in the past and that the presented numbers should not be considered as a realistic scientific resources evaluation but more 
as an educated optimistic guess for an upper limit resource estimate.  

The more speculative, not yet found but believed to exist, 
amount of inferred uranium (IR) resources are presented in Table 5.

In our previous report about the 2007 edition of the Red Book, we noted that the reported large increase of 0.4 million tons 
in the $<$ 40 dollars/kg IR category was very suspicious. 
It is thus interesting that this number is now about 1 million tons smaller than in 2007 edition. 
This change is even more remarkable as the resources assigned to this RAR and IR price category were reported with a continuous increase of up to 0.2 million tons 
in the previous Red Book editions from 2001 to 2005.  
Surprisingly,  the low-cost IR number dropped to 1/6 of its 2007 value in 2009, twice as strong as the corresponding drop in the low-cost RAR category.
      
{\small
\begin{table}[h]
\vspace{0.3cm}
\begin{center}
\begin{tabular}{|c|c|c|c|}
\hline

Red Book year      & IR  [tons]                                    & IR     [tons]                        & IR  [tons]\\
                                & less than 40 dollars/kg          & 40-80 dollars/kg             & 80-130 dollars/kg   \\
\hline
2001                    &  552 000               &    186 950                             & 225 150  \\
2003                    &  792 782               &    275 170                             & 320 868  \\
2005                    &  798 997               &    362 041                             & 285 126  \\
2007                    & 1 203 600               &   655 480                             & 272 200 \\
2009                    &   226 600               &   999 200                             & 653 300 \\

\hline

\end{tabular}\vspace{0.1cm}
\caption{The evolution of the not yet discovered but believed to exist IR uranium resources 
as given in the last five editions of the Red Book. Remarkable is the claim that 
the cheaper cost categories used to increase by ``large amounts" but as a result of the new cost evaluation has essentially 
disappeared and the amount has been roughly shifted to the higher cost categories. 
The new high cost category of 130-260 dollars/kg is reported with an amount of 422700 tons.} 
\end{center}
\end{table}
}

In their press release the authors of the Red Book explain these changes in terms of increased taxes combined with unexplained "increased mining costs".  In absence of explanations we are left with two big questions: (1) what is it, exactly, that accounts for mining costs more than doubling, and (2) why does this big increase in mining costs affect the different well understood RAR cost categories in a completely different way than in the corresponding IR categories.   

For completeness the evolution for the undiscovered prognosticated and speculative UPR and USR
resource categories with a given cost estimate is given in Table 6. Somehow surprisingly and despite the introduction of the 130-260 dollar/kg cost range, 
the amount of uranium in the UPR category has remained essentially unchanged but the one in the USR has decreased by almost 0.9 million tons.
This large change in the USR category comes almost entirely from Kazakhstan with a decrease from 0.5 to now 0.3 millions tons 
and from Russia,  where the entire estimated USR amount has been moved from a price tag of less than 130 dollar/kg to 
an unassigned cost range. Thus only by adding the $<$ 260 dollar/kg with the even more doubtful unassigned cost category 
the USR numbers have decreased from about 7.8 million tons in 2007 to 7.5 million tons in 2009. 

These perhaps bizarre but relatively minor changes for the least understood UPR and USR resource numbers, especially in comparison with the RAR and IR numbers, confirm that many of the Red Book data are not based on geological estimates.

{\small
\begin{table}[h]
\vspace{0.3cm}
\begin{center}
\begin{tabular}{|c|c|c|c|}
\hline

Red Book year      & UPR    [tons]                                     & UPR  [tons]                           & USR   [tons]                                      \\
                                & less than 80 dollars/kg          & 80-130 dollars/kg             & less than 130 dollars/kg          \\
\hline
2001                    & 1 480 000               &    852 000                             & 4 438 000  \\
2003                    & 1 474 600               &    779 900                             & 4 437 300  \\
2005                    & 1 700 100               &    818 700                             & 4 557 300  \\
2007                    & 1 946 200               &    822 800                             & 4 797 800 \\
2009                    & 1 701 500               &   1 113 300                           & 3 738 200 \\

\hline
\end{tabular}\vspace{0.1cm}
\caption{The evolution of the undiscovered prognosticated UPR and speculative USR uranium resources 
according to the past four Red Book editions. In comparison to the large relative changes in the IR data,  
the UPR numbers show an almost astonishing ``stability". The apparent large decrease in the USR with an assigned cost category 
is compensated by a large increase in the USR category with unassigned costs.  
The newly introduced cost category of 130-260 dollars/kg 
contains remarkable small numbers of 90200 tons (UPR) and 163 500 tons (USR).}
\end{center}
\end{table}
}

We now turn to some particular important countries for uranium mining and their reported resource changes. 

The newly introduced price category of 130-260 dollar/kg contains an overall estimated amount of 480 000 tons (RAR) 
and 422 000 tons (IR). 
It is remarkable that more than 85\% of the change in this cost category for RAR comes from just three countries --  USA (265 000 tons), Kazakhstan (78 000 tons), and the Ukraine (66 000 tons).  In comparison, the uranium-rich countries of Australia and Canada are said to be contributing just 3000 and 16 000 tons respectively to this high cost category.

Even if we ignore the fact that the USA did not report numbers for high-cost IR (130-260 dollar/kg), a similar imbalance among countries is found in this category.  Large numbers are reported for Kazakhstan (102 000 tons), Russia (86 000 tons), Ukraine (51 000 tons) and Canada (33 000 tons) while Australia is reported as a negligible 3000 tons -- in contrast to the very large changes reported by Australia in the lower-cost RAR and IR categories.  

And surprisingly in absence of detailed country reports the authors of the Red Book have estimated
relatively large amounts of 85 600 tons and 19 500 tons in this high-cost IR category for Denmark and Tanzania respectively 
-- literally unbelievable given than neither country reports numbers in any other category.  One can only wonder what method led to these estimates.

Table 7 and 8 show the RAR uranium resources for some countries in the $<$ 40 dollars/kg  
and 40-130 dollars/kg categories. As described above, the previous confident predictions for the $<$ 40 dollar/kg category have mostly been moved to the 40-130 dollar/kg category.  However, looking at the country-by-country situation, it is remarkable that almost the entire change can be explained with presumed mining cost explosions in Australia, Kazakhstan and Russia, while the claimed resource situation in Canada remained essentially unaffected.  

Concerning the not known, but believed to exist, IR uranium category, the numbers indicate that the hoped for amount in the 40-80 dollar/kg category 
is now dominated by Australia which claims about 45\% of the total estimated amount of about 1 million tons. 
This claim should be compared with their estimated 
fractions of about 7.5\% of the total of 650 000 tons in the 80-130 dollar/kg category and less than 1\% of the 420 000 tons in the 130-260 dollar/kg category.

People who argue that uranium is so plentiful that -- given high enough prices -- we will have hundred of years of supplies, need to explain the reported cost category data for RAR and IR uranium resources in Australia.  Specifically they need to explain why the RAR and IR numbers for Australia in the 40-80 dollars/kg are estimated to be 1.16 and 0.449 million tons respectively while the corresponding amounts for the 80-260 dollar/kg cost range, 0.016 and 0.051 million tons, are essentially negligible.

{\small
\begin{table}[h]
\vspace{0.1cm}
\begin{center}
\begin{tabular}{|c|c|c|c|}
\hline
country      & RAR (RB05)                             & RAR (RB07)                          & RAR (RB09)                              \\
                  &     40 dollars/kg  [tons]                          & 40 dollars/kg  [tons]  & 40 dollars/kg  [tons]                  \\
\hline

Australia                              &   701 000                             & 709 000                          & NA \\
Kazakhstan                          &   278 840                             & 235 500  &        14 600  \\
Namibia                               &   62 186                             & 56 000            &                0\\
Niger                                   &   172 866                             & 21 300                 &        17 000 \\
Russia                                 &   57 530                             & 47 500                &                0 \\
USA                                    &   NA                                    & NA               &            0 \\
sum                                   &     1 272 422                        &  1 069 300    &     31 600\\ 
\hline 
Canada                                &   287 200                               & 270 100             & 267 100\\
\hline
\end{tabular}\vspace{0.1cm}
\caption{Evolution of the low cost RAR uranium category for some countries important for uranium
mining. Note the very large declines shown for Australia, Kazakhstan, Namibia, and Russia between 2007 and 2009.}
\end{center}
\end{table}
}

{\small
\begin{table}[h]
%\vspace{0.1cm}
\begin{center}
\begin{tabular}{|c|c|c|c|}
\hline

country        & RAR (RB05)                                                            & RAR (RB07)                        & RAR (RB09)\\
                    &     40-130 dollars/kg  [tons]                          & 40-130 dollars/kg  [tons]              & 40-130 dollars/kg  [tons]  \\
\hline

Australia                  &   46 000                             & 16 000        & 1 176 000\\
Kazakhstan             &   235 057                             & 142 600     & 321 600\\
Namibia                  &   120 370                             & 120 000     & 157 000 \\
Niger                      &   7 600                             & 222 180         & 225 000    \\
Russia                   &   74 220                             & 124 900        & 181 400   \\
USA$^{*}$               &   342 000                             & 339 000       &  207 400   \\
sum                      &    825 247                         &   947 680      &  2 268 400\\
\hline
Canada                   &      58 000                               & 59 100    &  94 000 \\

\hline                 
\end{tabular}\vspace{0.1cm}
\caption{Evolution of the higher cost RAR uranium category for countries which claim to have 
a total of more than 100 000 tons of RAR resources on their territory and were large changes were reported. Especially remarkable 
are the changes from 2005 to 2009 for Australia, Kazakhstan, Niger and Russia. The observed increases are largely explained 
by the corresponding loss 
in the low cost category presented in Table 6.$^{*}$The USA did not report the $<$ 40 dollars/kg RAR
category separately and the amount in the $<$ 130 dollars/kg category is used.}
\end{center}
\end{table}
}

The reported RAR changes, see Tables 7 and 8,  are especially important for Australia, Kazakhstan, Russia, Niger and Namibia.
Large relative changes in the RAR cost categories are also reported from the ``former Soviet-Union" states Ukraine and from Uzbekistan.
These drastic changes can be compared with the ones from Canada which remain essentially unchanged.

It might be argued that contrary to the pretended Red Book accuracy, the reporting countries do not use the same methodology to estimate their 
resource data.
Some countries might take a serious attitude towards cost estimates for future uranium mining, while others just feel obliged to 
provide some rough estimates for possible upper limits to the biannual report. With the experience that mining costs have increased 
also because of the rising oil price and without better estimates, some large ``inflation-adjusted" cost estimate and the corresponding 
changes could be imagined as justified.

Concerning the question if and how the yearly uranium mining changes the RAR numbers, the data from  
Canada in the $<$ 40 dollar/kg 
cost category are interesting. Between 2005 and 2007 the numbers decreased by 17100 tons, an amount witch matches well with the 
amount extracted during 2005 and 2006. It follows that the operating mines in this country exploit dominantly the low cost 
category. Unfortunately, the uranium extracted in 2007 and 2008 came from the same 
uranium deposits, the reported resources did not decrease accordingly. The logical conclusion is that the 2009 numbers for Canada were not corrected 
for the extractions in 2007 and 2008. Unfortunately for the other countries with important uranium mining the 
RAR data have changed by much larger amounts, related to the large cost increase, and no indication about the treatment of past uranium extraction can 
be obtained. One might hope that the editors of the Red Book might eventually explain how uranium extraction in the different 
countries changes the resource data.   

For completeness Table 9 and 10 show the corresponding IR uranium resources for the same countries as above. 
As described above for like for the RAR categories, the previous estimates for the $<$ 40 dollar/kg category have mostly been moved to the 40-130 dollar/kg category. 
{\small
\begin{table}[h]
\vspace{0.1cm}
\begin{center}
\begin{tabular}{|c|c|c|c|}
\hline
country      & IR (RB05)                             & IR (RB07)                          & IR (RB09)                              \\
                  &     40 dollars/kg  [tons]                          & 40 dollars/kg  [tons]  & 40 dollars/kg  [tons]                  \\
\hline

Australia                              &   343 000                             & 487 000                          & NA \\
Kazakhstan                          &   129 252                              & 281 800  &   29 800       \\
Namibia                               &   61 192                             & 60 400            &                0\\
Niger                                   &        0                             & 12 900                 &                0 \\
Russia                                 &   21 572                             & 36 100                &                0 \\
sum                                   &     555 016                        &  878 200                &        29 800  \\ 
\hline 
Canada                                &    84 600                               & 82 300             &99 700\\
\hline
\end{tabular}\vspace{0.1cm}
\caption{Evolution of the low cost IR uranium category for some countries important for uranium
mining. Note that this IR category, after years of increase became essentially negligible for many countries in the 2009 edition.}
\end{center}
\end{table}
}

{\small
\begin{table}[h]
%\vspace{0.1cm}
\begin{center}
\begin{tabular}{|c|c|c|c|}
\hline

country        & IR (RB05)                                                            & IR (RB07)                        & IR (RB09)\\
                    &     40-130 dollars/kg  [tons]                          & 40-130 dollars/kg  [tons]              & 40-130 dollars/kg  [tons]  \\
\hline

Australia                  &   53 000                             & 31 000        & 497 000 \\
Kazakhstan             &   172 950                             & 157 400     & 285 800\\
Namibia                  &     38 611                             &   38 200     & 127 200 \\
Niger                      &     44 903                            & 18 000         &  30 900    \\
Russia                   &   19 080                             & 337 200        & 298 900   \\
sum                      &      328 544                         &   581 800      &  1 239 800\\
\hline
Canada                   &      14 000                               & 11 700    &  24 500 \\
\hline                 
\end{tabular}\vspace{0.1cm}
\caption{Evolution of the higher cost IR uranium category for some countries important for uranium mining. 
Essentially the entire loss in the low cost IR (Table 9) category has been absorbed into the higher cost categories. }
\end{center}
\end{table}
}

Again, as discussed in our analysis of the 2007 edition of the Red Book, a detailed examination of the country-by-country 2009 data do not justify the claim that conventional uranium resources increased by another 15\% in two years. Instead one is left with two conclusions: (1) that good data regarding conventional uranium resources does not exist and (2) that there is probably an agenda of some sort behind the data reported by most if not all countries.
Thus guesses about the possibility to fuel ``a bright future for nuclear fission energy" are currently only based on bad quality uranium data. 

All this could be summarized with the statement that the Red Book uranium country by country data are highly unreliable and nobody knows if 
the claimed amount of uranium is really exploitable. However, an alternative for a realistic estimate 
about the future of uranium mining and exploitable resource could be based on 
the experience with past, present and soon to be operated uranium mines. A more serious approach towards more reliable uranium data would thus 
be to analyze mine by mine and compare original estimates about the geological amount of exploitable resources with the real annual extraction profile. 
Furthermore financial extraction costs should be complemented with additional and realistic cost estimates for the entire mining life cycle. 
The total uranium extraction cost estimate should thus start from the amount needed to discover an interesting deposit and should quantify the amount of energy and other materials and their cost estimate during the exploration phase. Honest cost estimates would also include the amount of money lost due to the environmental 
damage during the exploration and its later repair.  We believe that such estimates are needed by all interested parties no matter what their attitude towards nuclear energy might be.

\section{Summary} 

Our analysis of the latest publicly available data from the large international and very pro nuclear organizations, the IAEA 
and the WNA, show that the 2009 and preliminary 2010 data regarding nuclear fission energy are still consistent with an average slow decline from the older reactors of perhaps 0.5\% per year. This situation is summarized by the following points:

\begin{itemize}
\item The percentage of electric energy provided by nuclear fission has declined from 18\% in 1993 to less than 14\% in 2009. The total worldwide production of nuclear power plants in 2009 -- 2560 TWhe -- was the lowest since at least 2005; it was about 1.6\% lower than 2008 production and almost 3.7\% lower than the record of 2658 TWhe set in 2006.  Preliminary data from the first 10 months of 2010 indicate that nuclear energy production in the OECD countries will increase back to the 2008 numbers. 

\item Today some 65 nuclear power plants with a capacity of more than 60 GWe 
are under construction worldwide. However, only 15\% of them are being constructed within the OECD countries, 
which currently host about 85\% of the existing nuclear reactors. 
Since something like one third of the older reactors in the OECD countries will reach retirement age during the next ten years, further declines in nuclear energy production in the OECD countries -- and especially in Europe -- seem inevitable."

\item The natural uranium equivalent required to operate the 374 GWe nuclear power plants plus the initial fueling of a few new reactors in 2010 has been estimated by the WNA to be about 
68 000 tons per year. This amount needs to be compared with the 51 000 tons extracted during 2009, a large increase compared to 
previous years, and with the 55 000 tons expected for 2010. 
This difference between requirements and mining, currently compensated by secondary reserves, needs to 
be reduced strongly during the coming years. 
However, the latest Red Book data about existing and soon to open uranium mines indicate that, even in the unlikely case that all projects 
will start operation according to plans, the uranium supply situation will 
remain problematic over the coming years. In fact based on the Red Book's data it is difficult to imagine how enough primary uranium can possibly be extracted to fuel all of the ambitious nuclear energy projects that are planned by China, India and Russia.
\end{itemize}

Concerning a first analysis of the just published 2009 edition of the Red Book on uranium resources we find that 
the data are at least as unconvincing as that of previous editions and do not justify claims of a 15\% increase in conventional uranium resources. 
Nevertheless, perhaps the two most remarkable observations from the new Red Book are that:
\begin{itemize} 
\item the best understood RAR and IR resources, with a price tag of less than 40 US dollars/Kg, have essentially disappeared and  
inconsistently reappeared in one or the other or both of the two-to-three-times higher-cost categories of RAR and IR;
\item the uranium mining boom in Kazakhstan is presented with a short lifetime. The presented mining capacity numbers indicate that the uranium mining in this country will be increased 
from 18 000 tons in 2010 to an extraction peak of 
28 000 tons during the years 2015-2020, followed by a steep decline to 14 000 tons by 2025 and to only 5000-6000 tons by 2035. 
\end{itemize}

\vspace{1.cm}
\noindent
{\bf \large Acknowledgments} \\
{\normalsize  \it{
Even though the views expressed in this paper are from the author alone, I would like to thank several colleagues, friends and many students 
who took the trouble to discuss the subject during the past few years with me. I would like to thank especially W. Tamblyn for many valuable suggestions and his careful reading of this paper draft. I would also like to thank Prof. F. Cellier for the many useful discussions about the subject and his encouragement to write this and the previous reports, ``The Future Of Nuclear Energy", for the Oil Drum.
}
}

%\newpage

\end{document}